\documentclass[12pt,letter]{article}

\usepackage{graphicx, epsfig, color}
\usepackage{amsmath,amssymb,hyperref}
\usepackage{subfigure}

\def\eslt{\not\!\!{E_T}}
\def\to{\rightarrow}

\def\bi{\begin{itemize}}
\def\ei{\end{itemize}}
\def\te{\tilde e}

\def\tu{\tilde u}
\def\sps1ap{SPS1a$^\prime$}
\def\c1p{C1$^\prime$}

\def\tb{\tilde b}
\def\tf{\tilde f}

\def\tst{\tilde t}
\def\ttau{\tilde \tau}

\def\tg{\tilde g}
\def\tnu{\tilde\nu}

\def\tw{\widetilde W}
\def\tz{\widetilde Z}
\def\alt{\lesssim}
\def\agt{\gtrsim}
\def\be{\begin{equation}}  
\def\ee{\end{equation}}  
\def\bea{\begin{eqnarray}}  
\def\eea{\end{eqnarray}}  
\def\beas{\begin{eqnarray*}}  
\def\eeas{\end{eqnarray*}}  
\newcommand\prd[3]{{\it Phys.\ Rev.\ }{\bf D #1} (#2) #3}

\newcommand\prl[3]{{\it Phys.\ Rev.\ Lett.\ }{\bf #1} (#2) #3}
\newcommand\plb[3]{{\it Phys.\ Lett.\ }{\bf B #1} (#2) #3}
\newcommand\jhep[3]{{\it J. High Energy Phys.\ }{\bf #1} (#2) #3}

\newcommand\npb[3]{{\it Nucl.\ Phys.\ }{\bf B #1} (#2) #3}

\newcommand\ptp[3]{{\it Prog.\ Theor.\ Phys.\ }{\bf #1} (#2) #3}

\newcommand{\hepph}[1]{hep-ph/#1}

\begin{document}
\begin{titlepage}
\begin{flushright}
OUHEP-161020\\
UH-511-1267-16
\end{flushright}

\vspace{0.5cm}
\begin{center}
{\Large \bf Natural generalized mirage mediation
}\\ 
\vspace{1.2cm} \renewcommand{\thefootnote}{\fnsymbol{footnote}}
{\large Howard Baer$^1$\footnote[1]{Email: baer@nhn.ou.edu }, 
Vernon Barger$^2$\footnote[2]{Email: barger@pheno.wisc.edu },
Hasan Serce$^1$\footnote[3]{Email: serce@ou.edu } and 
Xerxes Tata$^3$\footnote[3]{Email: tata@phys.hawaii.edu }
}\\ 
\vspace{1.2cm} \renewcommand{\thefootnote}{\arabic{footnote}}
{\it 
$^1$Department. of Physics and Astronomy,
University of Oklahoma, Norman, OK 73019, USA \\
}
{\it 
$^2$Department. of Physics,
University of Wisconsin, Madison, WI 53706, USA \\
}
{\it 
$^3$Department. of Physics,
University of Hawaii, Honolulu, HI 96822, USA \\
}

\end{center}

\vspace{0.5cm}
\begin{abstract}
\noindent

In the supersymmetric scenario known as mirage mediation (MM),
the soft SUSY breaking terms receive comparable anomaly-mediation and moduli-mediation
contributions leading to the phenomenon of {\it mirage unification}.
The simplest MM SUSY breaking models which are consistent with the
measured Higgs mass and sparticle mass constraints are strongly
disfavoured by fine-tuning considerations.
However, while MM makes robust predictions for gaugino masses, 
the scalar sector is quite sensitive to specific mechanisms for moduli stabilization and potential uplifting. 
We suggest here a broader setup of generalized mirage mediation (GMM), 
where heretofore discrete parameters are allowed as continuous to better 
parametrize these other schemes. 
We find that natural SUSY spectra consistent with both the measured value of $m_h$ 
as well as LHC lower bounds on superpartner masses are then possible.
We explicitly show that models generated from natural GMM
may be beyond the reach of even high-luminosity LHC searches. 
In such a case, the proposed International Linear $e^+e^-$ Collider (ILC) will be
required for natural SUSY discovery via higgsino pair production reactions. 
We also outline prospects for detection of higgsino-like WIMPs from natural GMM.

\vspace*{0.8cm}

\end{abstract}

\end{titlepage}

\section{Naturalness in Mirage Mediation}

Superstring theory yields a consistent quantum
theory of gravity and appears to have the required ingredients to
potentially unify all four forces of nature. However, in order
to gain predictivity, it is necessary to understand how the degeneracy
associated with the many flat directions in the space of scalar fields
(the moduli) is lifted to yield the true ground state, since many
quantities relevant for physics at low energy are determined by the
ground state values of these fields. The implementation of a class of
compactifications where the extra spatial dimensions are curled up to
small sizes with fluxes of additional fields trapped along these extra
dimensions was used by Kachru, Kallosh, Linde and Trivedi (KKLT)~\cite{kklt} to
construct models with a stable, calculable ground state with a
positive cosmological constant and broken supersymmetry. The KKLT toy
model is based on type-IIB superstrings including compactification with
fluxes to a Calabi-Yau orientifold. While the background fluxes serve
to stabilize the dilaton and the moduli that determine the shape of the
compact manifold, it is necessary to invoke a non-perturbative mechanism
such as gaugino condensation~\cite{inocon} on a $D7$ brane to stabilize the size of
the compact manifold. Finally, a non-supersymmetric anti-brane
($\overline{D3}$) was included in order to break supersymmmetry
completely and
obtain a de Sitter universe as required by observations. The resulting
low energy theory thus has no unwanted light moduli, has a broken
supersymmetry, and a positive cosmological constant.
The existence of these flux compactifications with stable calculable minima 
having many desired properties may be viewed as
a starting point for the program of discovering a string ground
state that may lead to a phenomenologically viable low energy theory of
SM particles and their superpartners, 
with $N=1$ supersymmetry softly broken just above the weak scale. 

The KKLT picture motivated several groups to analyze 
the structure of the soft SUSY breaking (SSB) terms in models based on 
a generalization of the KKLT set-up~\cite{choi}. 
The key observation is that because of the mass hierarchy,
\begin{equation}
m_{\rm moduli}\gg m_{3/2}\gg m_{\rm SUSY} ,
\label{eq:hierarchy}
\end{equation}
that develops in these models, the soft terms receive comparable
contributions from both modulus (gravity)~\cite{sugra} and anomaly
mediation of SUSY breaking~\cite{amsb}, with their relative size
parametrized by an additional parameter $\alpha$. Moreover, the
hierarchy (\ref{eq:hierarchy}) that leads to this mixed modulus-anomaly
mediated SUSY breaking (also known as mirage-mediation or MM as
discussed shortly) automatically alleviates phenomenological problems
from late decaying moduli and gravitinos that could disrupt, for
instance, the predictions of light element abundances from Big Bang
nucleosynthesis. Upon integrating out the heavy dilaton field and the
shape moduli, one is left with an effective broken supergravity theory
of the observable sector fields denoted by $\hat{Q}$ and the size
modulus field $\hat{T}$. The K\"ahler potential depends on the location
of matter and Higgs superfields in the extra dimensions via their
modular weights $n_i = 0 \ (1)$ for matter fields located on $D7$ ($D3$)
branes, or $n_i=1/2$ for chiral multiplets on brane intersections, while
the gauge kinetic function $f_a={\hat{T}}^{l_a}$, where $a$ labels the
gauge group, is determined by the corresponding location of the gauge
supermultiplets, since the power $l_a= 1 \ (0)$ for gauge fields on $D7$
($D3$) branes~\cite{choi3}.

Within the MM model, the SSB gaugino mass parameters, trilinear SSB
parameters and sfermion mass parameters, all renormalized just below the
unification scale (taken to be $Q=M_{\rm GUT}$), are given by,
\begin{eqnarray}
M_a&=& M_s\left( l_a \alpha +b_a g_a^2\right),\label{eq:M}\\
A_{ijk}&=& M_s \left( -a_{ijk}\alpha +\gamma_i +\gamma_j +\gamma_k\right),
\label{eq:A}\\
m_i^2 &=& M_s^2\left( c_i\alpha^2 +4\alpha \xi_i -
\dot{\gamma}_i\right) ,\label{eq:m2}
\end{eqnarray}
where $M_s\equiv\frac{m_{3/2}}{16\pi^2}$,
$b_a$ are the gauge $\beta$ function coefficients for gauge group $a$ and 
$g_a$ are the corresponding gauge couplings. The coefficients that
appear in (\ref{eq:M})--(\ref{eq:m2}) are given by
$c_i =1-n_i$, $a_{ijk}=3-n_i-n_j-n_k$ and
$\xi_i=\sum_{j,k}a_{ijk}{y_{ijk}^2 \over 4} - \sum_a l_a g_a^2
C_2^a(f_i).$ 
Finally, $y_{ijk}$ are the superpotential Yukawa couplings,
$C_2^a$ is the quadratic Casimir for the a$^{th}$ gauge group
corresponding to the representation to which the sfermion $\tf_i$ belongs,
$\gamma_i$ is the anomalous dimension and
$\dot{\gamma}_i =8\pi^2\frac{\partial\gamma_i}{\partial \log\mu}$.
Expressions for the last two quantities involving the 
anomalous dimensions can be found in the Appendix of Ref.~\cite{flm,cjko}.

The MM model is then specified by the parameters
\begin{equation}
\ m_{3/2},\ \alpha ,\ \tan\beta ,\ sign(\mu ),\ n_i,\ l_a. 
\label{eq:par1}
\end{equation}
The mass scale for the SSB parameters is dictated by the gravitino mass
$m_{3/2}$. The phenomenological parameter $\alpha$, which could be of
either sign, determines the relative contributions of anomaly mediation
and gravity mediation to the soft terms, and is expected to be $|\alpha|
\sim {\cal O}(1)$. Grand Unification implies matter particles within
the same GUT multiplet have common modular weights, and that the $l_a$
are universal. We will assume here that all $l_a=1$ and, for
simplicity, there is a common modular weight for all matter scalars
$c_m$ but we will allow for different modular weights $c_{H_u}$ and
$c_{H_d}$ for each of the two Higgs doublets of the MSSM. Such choices
for the scalar field modular weights are motivated for instance by
$SO(10)$ SUSY GUT models where the MSSM Higgs doublets may live in different
$\bf 10$-dimensional Higgs reps.

Various aspects of MM phenomenology have been examined in
Refs.~\cite{choi3,flm,eyy,kn,bptw}. The universality of the $l_a$ leads
to the phenomenon of {\it mirage unification}~\cite{choi3,flm} of
gaugino mass parameters (and also corresponding matter scalar mass
parameters of first and second generation sfermions whose Yukawa
couplings are negligible). Here, for reasons that will become clear
later, we focus on the gaugino mass parameters $M_i$: when
extrapolated to high energies using one loop renormalization group
equations (RGEs), these will unify at a scale $Q=\mu_{\rm mir} \not= M_{\rm GUT}$, 
where $M_{GUT}$ is the unification scale for gauge couplings. 
Indeed, the observation of gaugino mass unification at the mirage unification scale,
\be
\mu_{\rm mir}=M_{\rm GUT}\ e^{-8\pi^2/\alpha } ,
\label{eq:mir}
\ee
is the smoking gun of such a scenario~\cite{robust}. 
If $\alpha < 0$, then $\mu_{\rm mir}> M_{\rm GUT}$ and one finds {\it virtual} 
mirage unification at super-GUT energy scales. We stress that there is no physical threshold
at $Q= \mu_{\rm mir}$, and the evolution can be continued to $Q=M_{\rm GUT}$ 
where the gaugino mass parameters would take on the values close
to (\ref{eq:M}). The determination of the mirage unification scale
also determines $\alpha$, the parameter that governs the relative
moduli- versus anomali-mediation contribution to the soft SUSY breaking
terms. Once $\alpha$ is known, then further extrapolation of the
gaugino masses to $Q=m_{\rm GUT}$ allows for a determination of the {\it gravitino mass} $m_{3/2}$.

Alas, this attractive MM scenario has recently been confronted by the
twin constraints of LHC searches on the one hand and a clarified
understanding of SUSY naturalness on the other. One important LHC constraint
comes from the new-found Higgs mass $m_h\simeq 125$ GeV which in the
context of the MSSM requires highly mixed TeV-scale
top-squarks~\cite{h125}. The other LHC constraint is that the gluino
mass, based on LHC13 searches with $\sim 10$ fb$^{-1}$ of data, require
$m_{\tg}\agt 1.9$ TeV (within the context of various simplified
models)~\cite{atlas_s}. 

For the case of naturalness, it has been emphasized~\cite{Baer:2013gva,dew,mt}
that previous studies-- that lead to the conclusion that naturalness
requires light top squarks-- neglect the fact that one must evaluate the
sensitivity of $m_h$ or $m_Z$ only with respect to the {\it independent}
parameters of the theory, as embodied for instance in the 
%
frequently used EENZ/BG measure~\cite{bg}, 
$\Delta_{\rm BG}\equiv max_i|\frac{\partial\log m_Z^2}{\partial \log p_i}|$. 
Here $i$ labels the various independent, fundamental parameters $p_i$ 
of the theory.
Historically, this measure has been applied to multi-soft-parameter effective 
SUSY theories where the additional parameters are introduced to 
parametrize our ignorance of the source of soft terms. 
However, in any more fundamental theory the various soft terms are derived 
in terms of more fundamental entities, such as the gravitino mass 
in gravity mediation~\cite{soft}, or via Eq. (\ref{eq:M})-(\ref{eq:m2}) for
mirage-mediation.
In this case, the soft-SUSY breaking parameters are {\it correlated} 
and not independent: then, neglecting these correlations will lead to an
{\it over-estimate} of the fine-tuning in these 
theories~\cite{Baer:2013gva,dew,mt}. 
In MM, where $\alpha$ takes on a pre-determined value, 
the soft parameters are all determined by $m_{3/2}$ and 
$\Delta_{\rm BG}$ reduces to the model-independent 
electroweak measure $\Delta_{\rm EW}$.\footnote{More
generally, we advocate the use of $\Delta_{\rm EW}$ in the discussion of
naturalness of models with a given superpartner spectrum since
discarding any high scale model with a (seemingly) large value of
$\Delta_{\rm BG}$ and a low value of $\Delta_{\rm EW}$ may be 
prematurely discarding an effective theory because 
(unincorporated) correlations among the high scale parameters 
could well lower the value of $\Delta_{\rm BG}$ 
all the way to $\Delta_{\rm EW}$: {\it das Kind mit dem Bade aussch\"utten!}}

The electroweak fine-tuning parameter~\cite{ltr,rns}, $\Delta_{\rm EW}$,
is a measure of the degree of cancellation between various contributions
on the right-hand-side (RHS) in the well-known expression for the $Z$ mass:
\be \frac{m_Z^2}{2} = \frac{m_{H_d}^2 +
\Sigma_d^d -(m_{H_u}^2+\Sigma_u^u)\tan^2\beta}{\tan^2\beta -1} -\mu^2
\simeq  -m_{H_u}^2-\Sigma_u^u-\mu^2 
\label{eq:mzs}
\ee 
which results from the minimization of the Higgs potential in the MSSM.
Here, $\tan\beta =v_u/v_d$ is the ratio of Higgs field
vacuum-expectation-values and the $\Sigma_u^u$ and $\Sigma_d^d$
contain an assortment of radiative corrections, the largest of which
typically arise from the top squarks. Expressions for the $\Sigma_u^u$
and $\Sigma_d^d$ are given in the Appendix of Ref.~\cite{rns}. 
If the RHS terms in Eq.~(\ref{eq:mzs}) are individually
comparable to $m_Z^2/2$, then no unnatural fine-tunings are required to
generate $m_Z=91.2$ GeV. $\Delta_{\rm EW}$ is defined to be the largest
of these terms, scaled by $m_Z^2/2$. Clearly, low electroweak
fine-tuning requires that $\mu$ be close to $m_Z^2$ and that $m_{H_u}^2$
be radiatively driven to {\it small} negative values close to the weak scale. 
This scenario has been dubbed radiatively-driven natural supersymmetry or RNS~\cite{ltr,rns}.
 
The main requirements for low electroweak fine-tuning ($\Delta_{\rm
EW}\alt 30$) \footnote{ The onset of fine-tuning for $\Delta_{\rm EW}\agt
30$ is visually displayed in Ref.~\cite{upper}.} are the following.
\bi
\item $|\mu |\sim 100-300$ GeV~\cite{mugev}
where $\mu \agt 100$ GeV is required to accommodate LEP2 limits 
from chargino pair production searches.
\item $m_{H_u}^2$ is driven radiatively to small, and not large,
negative values at the weak scale~\cite{ltr,rns}.
\item The top squark contributions to the radiative corrections
$\Sigma_u^u(\tst_{1,2})$ are minimized for TeV-scale highly mixed top
squarks~\cite{ltr}. This latter condition also lifts the Higgs mass to
$m_h\sim 125$ GeV. For $\Delta_{\rm EW}\alt 30$, the lighter top
squarks are bounded by $m_{\tst_1}\alt 3$ TeV~\cite{rns,upper}.
\item The gluino mass, which feeds into the stop masses and hence the $\Sigma_u^u (\tst_{1,2})$, 
is bounded by $m_{\tg}\alt 4$ TeV~\cite{rns,upper}.
\ei

Detailed scans over MM parameter space for various choices of matter and
Higgs field modular weights found all models consistent with LHC8
sparticle and Higgs mass constraints were in fact highly fine-tuned with
$\Delta_{\rm EW}> 100$ (for a summary, see Fig. 13 of Ref.~\cite{dew}).
This means these models give a poor prediction for the weak scale as
typified by $m_{\rm weak}\sim m_{W,Z,h}\sim 100$ GeV, {\it i.e.} the weak
scale of 100 GeV is only generated by excessive fine-tuning of the $\mu$
parameter.
One may thus ask: are mirage mediation models on their way to the
dustbin of failed SUSY models?\footnote{The models of deflected mirage
mediation~\cite{dmm} which combine gauge-, moduli- and anomaly-mediation, 
still seem viable and may allow for naturalness~\cite{todd}.}
\footnote{A phenomenological AMSB model has been proposed
which can reconcile $(g-2)_\mu$ with the value 
$m_h\simeq 125$ GeV\cite{Chowdhury:2015rja}.}

\section{Natural Generalized Mirage Mediation}

The evident failure of naturalness in MM mentioned at the end of the
last section leads us to re-examine the phenomenological implications of
moving from discrete choices of the parameters $a_{ijk}$ and $c_i$ in
Eqs.~(\ref{eq:A}) and~(\ref{eq:m2}) to a continuous range, and also to
allow $c_i$ values greater than 1. While the discrete parameter choices
occur in a wide range of KKLT-type compactifications (for some
discussion, see Ref.~\cite{choi_sax}), a continuous range of these
parameters may be expected if one allows for more generic methods of
moduli stabilization and potential uplifting.  For instance, if the soft
terms scan as in the string landscape picture, then their
moduli-mediated contributions may be expected to be parametrized by a
continuous value.  For models which generate a small $\mu$ term $\sim
100$ GeV from multi-TeV soft terms, such as radiative Peccei-Quinn
breaking~\cite{radpq}, it has been suggested that the statistical pull by
the landscape towards large soft terms, coupled with the anthropic
requirement of $m_{\rm weak}\sim 100$ GeV, acts as an attractor towards
natural SUSY soft term boundary conditions~\cite{landscape}.

Note that the
phenomenological modification that we suggest will not affect the result
(\ref{eq:M}) for gaugino mass parameters, which has been stressed~\cite{robust} 
to be the most robust prediction of the MM mechanism. In
this paper, we allow for the more {\it general} mirage mediation (GMM)
parameters, thus adopting a parameter space given by 
\be 
m_{3/2},\ \alpha,\ \tan\beta ,\ a_3,\ c_m,\ c_{H_u},\ c_{H_d} \ \ \ (GMM), 
\ee
where $a_3$ is short for $a_{Q_3H_uU_3}$. The independent values of
$c_{H_u}$ and $c_{H_d}$ which set the moduli-mediated contribution to
the soft Higgs mass terms may conveniently be traded for weak scale
values of $\mu$ and $m_A$ as is done in the two-parameter non-universal
Higgs model~\cite{nuhm2}: 
\be 
m_{3/2},\ \alpha,\ \tan\beta ,\ a_3,\ c_m,\ \mu,\ m_A \ \ \ (GMM^\prime ). 
\ee 
This trick allows for more
direct exploration of natural SUSY parameter space which requires
$\mu\sim 100-300$ GeV.

\begin{figure}[tbp]
\begin{center}
\includegraphics[height=0.4\textheight]{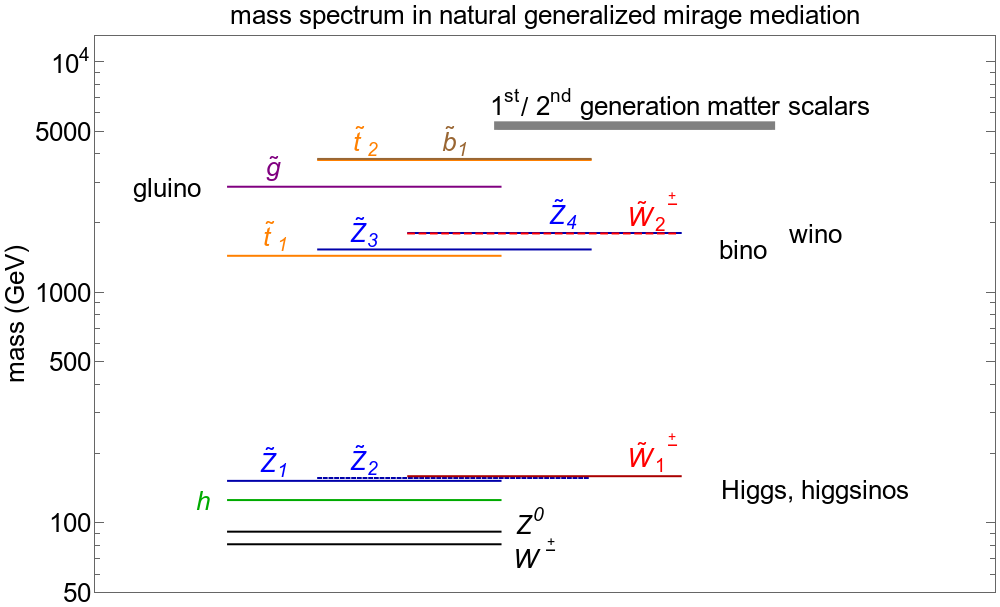}
\caption{A typical superparticle mass spectrum generated from 
natural generalized mirage mediation (nGMM) as in Table \ref{tab:bm}.
\label{fig:spect}}
\end{center}
\end{figure}
In Fig.~\ref{fig:spect}, we show the SUSY spectrum generated from one
such parameter space point in the {\it natural} GMM model (nGMM), with
the corresponding data shown in Table~\ref{tab:bm}. This benchmark point
was generated using the Isajet/Isasugra computer code~\cite{isajet} with
non-universal soft term inputs. The specific input parameters are
$m_{3/2}=75$ TeV, $\alpha =4$, $\tan\beta =10$, $a_3=5.1$, $c_m=6.9$ and
with $\mu =150$ GeV and $m_A=2$ TeV. The latter two choices end up
corresponding to $c_{H_u}=11.77$ and $c_{H_d}=1.15$. From
Table~\ref{tab:bm}, we see the gluino mass is $m_{\tg}=2841$ GeV, which
is just beyond the $5\sigma$ projected reach of HL-LHC with
$\sqrt{s}=14$ TeV and $3000$ fb$^{-1}$ of integrated
luminosity~\cite{hllhc}, at least without tagged $b$s to further enhance
the signal. 
The Higgs mass $m_h=125.3$ GeV agrees well with 
measurements from LHC.
The squarks and sleptons of the first/second generation lie in the 5 TeV
range while third generation squarks can be lighter, with
$m_{\tst_1}\simeq 1537$ GeV. This latter value appears beyond the reach
of HL-LHC where a 95\% exclusion reach with 3000 fb$^{-1}$ extends out
to $m_{\tst_1}\sim 1100$ GeV for $m_{\tz_1}\sim 100$
GeV~\cite{hl_atlas}. Note that this benchmark point has $\Delta_{\rm EW}
=18.6$ and so relatively low electroweak fine-tuning. A high scale
theory with $\alpha=4$ which led to the assumed values of $c_i$ and
$a_3$ would have $\Delta_{\rm BG} \simeq 18.6$ and would not be
fine-tuned. 

%
\begin{table}\centering
\begin{tabular}{lc}
\hline
parameter & nGMM \\
\hline
$m_{3/2}$      & 75000  \\
$\alpha$       & 4  \\
$\tan\beta$    & 10  \\
$c_{H_u}$      & 11.77  \\
$c_{H_d}$      & 1.15  \\
$c_m$          & 6.9  \\
\hline
$\mu$          & 150   \\
$m_A$          & 2000  \\
\hline
$m_{\tg}$   & 2841.4  \\
$m_{\tu_L}$ & 5242.1  \\
$m_{\tu_R}$ & 5382.3  \\
$m_{\te_R}$ & 4804.6  \\
$m_{\tst_1}$& 1537.3  \\
$m_{\tst_2}$& 3887.5  \\
$m_{\tb_1}$ & 3927.4  \\
$m_{\tb_2}$ & 5223.9  \\
$m_{\ttau_1}$ & 4776.5  \\
$m_{\ttau_2}$ & 5157.5  \\
$m_{\tnu_{\tau}}$ & 5168.5  \\
$m_{\tw_2}$ & 1800.2  \\
$m_{\tw_1}$ & 158.8  \\
$m_{\tz_4}$ & 1809.8  \\ 
$m_{\tz_3}$ & 1553.7  \\ 
$m_{\tz_2}$ & 155.9  \\ 
$m_{\tz_1}$ & 151.6  \\ 
$m_h$       & 125.3  \\ 
\hline
$\Omega_{\tz_1}^{std}h^2$ & 0.005  \\
$BF(b\to s\gamma)\times 10^4$ & $3.1$  \\
$BF(B_s\to \mu^+\mu^-)\times 10^9$ & $3.8$ \\
$\sigma^{SI}(\tz_1, p)$ (pb) & $2.8\times 10^{-10}$  \\
$\sigma^{SD}(\tz_1 p)$ (pb) & $9.1\times 10^{-6}$  \\
$\langle\sigma v\rangle |_{v\to 0}$  (cm$^3$/sec)  & $3.1\times 10^{-25}$ \\
$\Delta_{\rm EW}$ & 18.6 \\
\hline
\end{tabular}
\caption{Input parameters and masses in~GeV units
for a natural generalized mirage mediation SUSY benchmark point
with $m_t=173.2$ GeV and $a_3=5.1$.
}
\label{tab:bm}
\end{table}
\begin{figure}[tbp]
\begin{center}
\includegraphics[height=0.39\textheight]{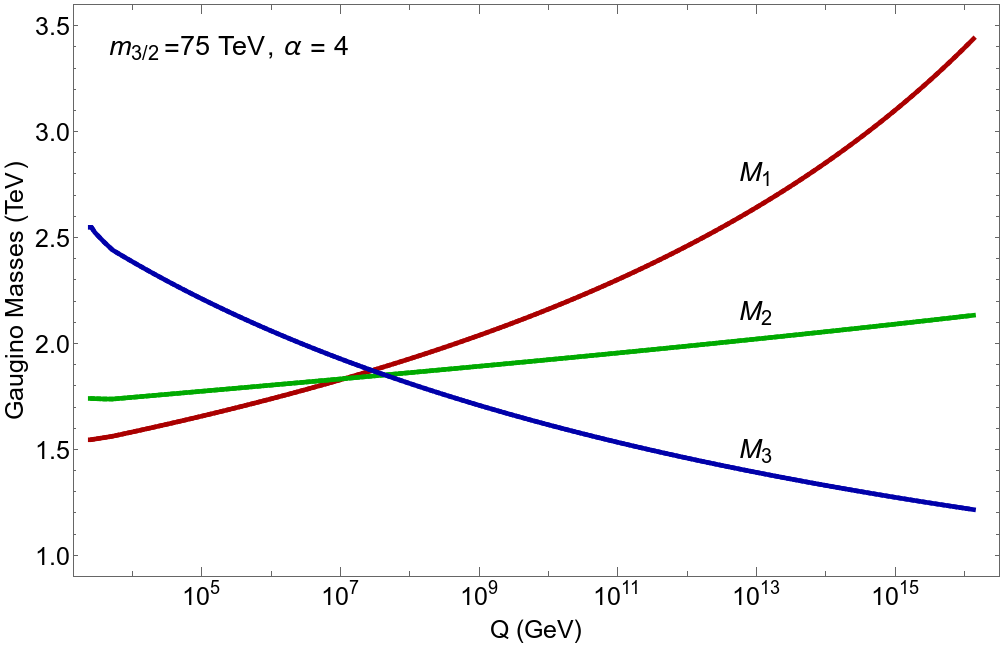}
\caption{Evolution of gaugino masses from the nGMM benchmark point with 
$m_{3/2}=75$ TeV, $\alpha =4$.
\label{fig:inos}}
\end{center}
\end{figure}

In Fig.~\ref{fig:inos}, we show the running of the three gaugino masses
for the nGMM benchmark model. In this case, we see the most robust
feature of GMM: the celebrated mirage unification of gaugino masses at
the intermediate scale $\mu_{mir}\sim 10^{7.5}$ GeV consistent with
$\alpha =4$, as can be seen from Eq.~(\ref{eq:mir}). 

\begin{figure}[tbp]
\begin{center}
\includegraphics[height=0.39\textheight]{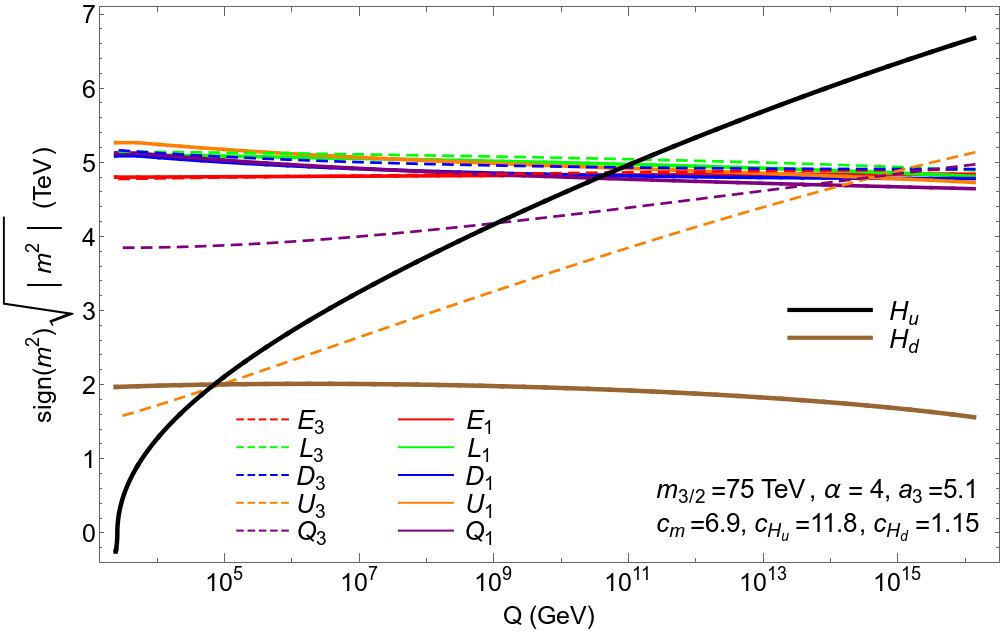}
\caption{Plot of running scalar masses from the nGMM benchmark point
with $m_{3/2}=75$ TeV, $\alpha =4$, $\tan\beta =10$ and $c_m=6.9$,
$a_3=5.1$ with $c_{H_u}=11.77$ and $c_{H_d}=1.15$ (corresponding to $\mu
=150$ GeV and $m_A=2$ TeV at the weak scale)
\label{fig:scalars}}
\end{center}
\end{figure}
In Fig.~\ref{fig:scalars}, we show the renormalization group evolution
of the various scalar soft mass terms for the nGMM benchmark
model. First/second generation matter scalar mass parameters remain
close to 5 TeV. Unlike for the model with a common modular weight for
the two Higgs doublets, these do not unify at $Q=\mu_{\rm mir}$ because
for the nGMM model, the hypercharge $D$-term contribution to the
evolution no longer vanishes. In contrast, third generation and Higgs
mass square parameters evolve considerably more because of large Yukawa
interactions. In particular, $m_{U_3}^2$, runs to much lower values
$\sim 1.5$ TeV at the weak scale. The up-Higgs soft mass $m_{H_u}^2$
begins about 20\% higher in value than matter scalar masses at $Q=M_{\rm GUT}$, 
but then evolves to small negative values at the weak scale, so
that the requirement for electroweak naturalness, $|m_{H_u}^2| \sim m_Z^2$ 
is satisfied.
The soft term $m_{H_d}^2$ which sets the heavy Higgs mass scale can be
adjusted up or down with not-to-much cost to naturalness $\Delta_{\rm EW}$. 
We remark here that because the matter scalars are essentially
decoupled, our spectra for phenomenological purposes is similar to
what may be derived from the NUHM2 model but with gaugino mass parameters 
fixed by the MM values rather than by universality. 

\begin{figure}[tbp]
\begin{center}
\includegraphics[height=0.4\textheight]{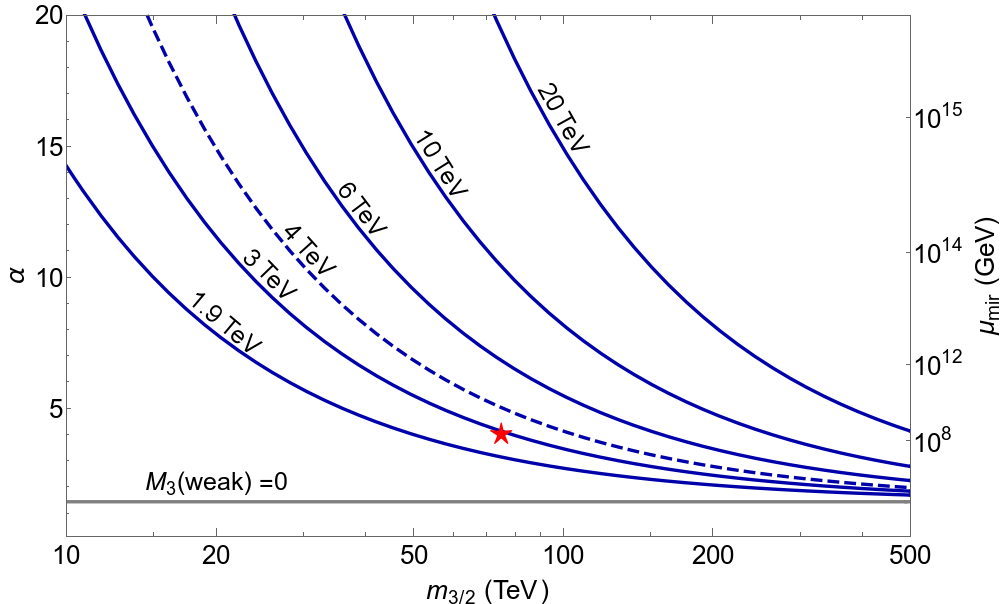}
\caption{Contours of $M_3(weak)$ in the $m_{3/2}$ vs. $\alpha$ plane of
  the nGMM model with other parameters as fixed in Table
  \ref{tab:bm} . The region below $M_3\sim 1.9$ is roughly excluded by
  LHC gluino pair searches. The location of our benchmark point is shown
  with a red star. 
  The region below the dashed $m_{\tg}=4$ TeV contour has the capacity to be natural.
  On the right side, some corresponding values of $\mu_{mir}$ are shown. 
\label{fig:M3}
}
\end{center}
\end{figure}
In Fig.~\ref{fig:M3}, we show a larger set of GMM parameter space by
contours of gaugino mass $M_3(weak)$ in the $m_{3/2}$ vs. $\alpha$ 
plane. At tree level, then $m_{\tg}\sim M_3(weak)$. Thus, the region below 
$M_3(weak)\alt 1.9$ TeV is excluded by LHC13 gluino pair searches.
The location of our benchmark point is noted with a red star. 
The region below the dashed $m_{\tg}=4$ TeV contour has the capacity to be natural.
On the right side, some corresponding values of $\mu_{mir}$ are shown.

\section{Consequences for Colliders}

\subsection{LHC}

It has been pointed out in Ref.~\cite{lhc2} that in natural SUSY models such as RNS
with gaugino mass unification, additional signatures for SUSY with light
higgsinos are present at the LHC even though gluinos and also top
squarks may be too heavy to be detectable. The first of these, labeled
same-sign diboson production~\cite{ssdb} (SSdB), arises from wino pair
production $pp\to\tw_4^\pm\tz_4$ where, for instance, $\tw_2^+\to
W^+\tz_{1,2}$ while $\tz_4\to W^+\tw_1^-$. This leads to a robust $W^\pm
W^\pm +\eslt$ signature consisting of two acollinear same-sign dilepton
+ $\eslt$ events with jet activity only from QCD radiation. These event
topologies have very low backgrounds. For large integrated luminosity
$\sim 300-3000$ fb$^{-1}$-- anticipated at the high luminosity LHC -- 
this channel yields the greatest LHC14 reach. 

A second robust signature expected in RNS-type models is higgsino pair
production $\tz_1\tz_2j$ in association with a hard monojet from QCD
radiation, followed by $\tz_2\to\tz_1 \ell^+\ell^-$ decay. The leptons
in the OS/SF pair emerging from $\tz_2$ decay are quite soft (due to the
small $m_{\tz_2}-m_{\tz_1}\sim 10-20$ GeV mass gap expected in models
with universal gaugino masses) and would frequently fail detector
trigger requirements. However, the hard ISR jet or the associated large
$\eslt$ could serve as a trigger. After suitable cuts, it appears this
signature gives a good reach in the $\mu$ direction of the
$\mu-m_{1/2}$ parameter plane of the model. The calculations of
Ref.~\cite{lhc2} indicate that essentially all of the RNS parameter
space with $\Delta_{\rm EW}\le 30$ is covered by these two channels assuming
$\sim 3000$ fb$^{-1}$ of integrated luminosity at LHC14.

In contrast, for the nGMM model, both these signatures appear much more
challenging for LHC SUSY searches. The reason is the compressed spectrum
of gauginos which occurs in nGMM. In NUHM2 with gaugino mass
unification at $Q=M_{\rm GUT}$, then the weak scale gauginos after RG
running are expected to occur in a ratio $M_3:M_2:M_1\sim 7:2:1$. 
Naturalness considerations 
require gluinos not much heavier than $\sim 4$ TeV 
in NUHM2 for $\Delta_{\rm EW}<30$~\cite{rns,upper}; if they do become
heavy, they increase the top-squark masses which increases the
$\Sigma_u^u(\tst_{1,2})$ contributions so that again one must fine-tune
against these contributions. This naturalness condition, together with
gaugino mass universality,
then guarantees
that the winos are almost always accessible to LHC14 searches for NUHM2
if $\Delta_{\rm EW} \leq 30$. Also, in this case, the $\tz_2-\tz_1$ mass gap 
is always larger than $\sim 10$~GeV.
In contrast, compressed gaugino spectra with $M_1 \sim M_2 \sim M_3$
at an intermediate scale are the hallmark of MM models
with a low $\alpha$ and concomitantly low mirage unification scale. 
This means that-- with $m_{\tg}\sim 3-4$ TeV--  wino pairs
(with mass $m(wino)\sim m_{\tg}$) may well be too heavy 
to be produced at detectable rates at LHC14. 
Moreover, these larger values of $M_1$ and $M_2$ from nGMM result in
an even more compressed spectrum of neutral higgsinos, as exemplified
by the benchmark in Table \ref{tab:bm} for which
the mass gap $m_{\tz_2}-m_{\tz_1}\sim 5.4$ GeV. 
Such a small mass gap makes the LHC monojet plus soft dilepton search
much more difficult -- in fact, in a recent CMS search for this
channel~\cite{cms}, they indeed required $m(\ell^+\ell^- )>4$ GeV 
to stay away from the $J/\psi$ and $\gamma^*$ poles with a cut around 
$9-10.5$ GeV to stay away from the $\Upsilon$ pole. 
Such cuts would veto much of the signal region expected from our nGMM benchmark.

\subsection{Linear electron-positron colliders}

In Ref.~\cite{kklt2}, a variety of measurements were proposed for MM models 
at the LHC and International Linear $e^+e^-$ Collider or ILC which could
determine the modular weights associated with matter scalars and 
measure the relative moduli-/anomaly-mediated contributions 
to the soft terms and the gravitino mass $m_{3/2}$. The ILC would initially be
operating with $\sqrt{s}=0.5$ TeV but is upgradable to 1 TeV.
In Ref.~\cite{ilc}, it was pointed out that for SUSY models with 
radiatively-driven naturalness, the ILC
would be a {\it higgsino factory} for $\sqrt{s}>2m(higgsino)\sim 2\mu$. 
The two reactions $e^+e^-\to\tw_1^+\tw_1^-$ and $\tz_1\tz_2$ 
occur at rates comparable to muon pair production once the 
kinematic production threshold is passed. Moreover,
the higgsino pair production cross section exceeds that for Higgs boson
production unless higgsino poduction is kinematically suppressed.  In
Ref.~\cite{ilc}, it was shown that the clean environment of ILC detector
events and the adjustable beam energy and polarization can easily allow
for both discovery as well as a suite of precision measurements, at
least for $\tz_1-\tz_2$ mass gaps expected in the RNS framework with
$\Delta_{\rm EW} < 30$. Direct measurement of the $E(\ell^+\ell^- )$ and
$m(\ell^+\ell^- )$ distributions from $\tz_1\tz_2$ production followed
by $\tz_2\to\tz_1\ell^+\ell^-$ decay allows for measurement of
$m_{\tz_2}$ and $m_{\tz_1}$ to sub-percent precision~\cite{ilc,
ilcgroup}. Measurement of the $E(jj)$ and $m(jj)$ distributions from
$\tw_1\tw_1\to (q\bar{q}'\tz_1)+(\ell\nu_\ell\tz_1)$ production allow
for sub-percent measurements of $m_{\tw_1}$ and $m_{\tz_1}$ if the mass
gap is sufficiently large. Moreover, the mass gaps are sensitive to
$\tan\beta $ and gaugino masses $M_1$ and $M_2$. In the RNS case with
$m_{\tz_2}-m_{\tz_1}\sim 20$~GeV, it was shown that the gaugino mass
parameters can be extracted with a precision of $5-10\%$, and
examination of the more difficult case of the 10~GeV mass gap is in
progress~\cite{ilcgroup}. Clearly, prospects for the detection of the
higgsinos of nGMM models (where the mass gap is even smaller) and
corresponding measurements of gaugino masses will be even more
challenging but worthy of investigation.\footnote{The nGMM model is not
the only scenario with a compressed higgsino spectrum and very heavy
gauginos that suggests that the ILC could be a discovery machine.  If
the vacuum-expectation-value of the auxiliary field that breaks
supersymmetry transforms as a {\bf 75} dimensional representation of
$SU(5)$ (rather than a singlet as is usually assumed), 
the resulting non-universal pattern of GUT scale gaugino masses leads to
$M_3:M_2:M_1 = 6,6,-5$ at the weak scale,
so winos and binos would be even heavier than for our nGMM case study,
and the higgsinos even more compressed. Such a scenario would be even
more challenging to detect.} A positive outcome would mean
that the ILC would be a {\em discovery machine} for a scenario that
would likely be beyond the reach of even a high luminosity LHC. We
emphasize that if the extraction of gaugino masses turns out to be
feasible, then extrapolation of these masses via RGEs to high energies
would indicate mirage unification and allow extraction of the parameter
$\alpha$, and also the associated gravitino mass $m_{3/2}$.

\section{WIMP signals from nGMM}

We now turn to prospects for dark matter detection 
in the natural generalized mirage mediation scenario. Since
electroweak naturalness requires a low $\mu$ parameter, $\mu\sim 100-300$
GeV, the LSP is expected to be mainly higgsino-like with a
non-negligible gaugino component. However, comparing nGMM to natural models
with gaugino mass unification like RNS, it is clear that for nGMM, the
electroweak gauginos are much heavier because the gaugino spectrum is 
more compressed.
As a result, both $\tz_1$ and $\tz_2$ are considerably more
higgsino-like than in RNS, and further, the inter-higgsino mass gaps are
also smaller. This, in turn, means the higgsino co-annihilation rate is
enhancd in nGMM relative to RNS. Consequently the thermally-produced
higgsino-like neutralino abundance can be as low as
$\Omega_{\tz_1}^{TP}h^2\sim 0.12/40$, {\it i.e.} thermally produced
higgsinos make up just a few percent of the observed DM, an even lower
relic abundance than in natural NUHM2 models.  The possibility that the
deficit in dark matter abundance is made up by non-thermal processes
such as moduli production and subsequent decay to higgsinos is excluded
as we will see below.
In Ref.~\cite{Bae:2013bva} it is
suggested that if one insists on naturalness in the electroweak sector,
one ought to have naturalness in the QCD sector: this brings into the
discussion axion superfields, mixed axion-higgsino dark matter and
production of neutralinos via axino/saxion production and decay. In this
latter case, then axions may make up the bulk of dark matter with only a
small fraction of the abundance consisting of higgsino-like WIMPs.

\begin{figure}[tbp]
\begin{center}
\subfigure[SI direct detection rate]
{\includegraphics[width=0.7\textwidth]{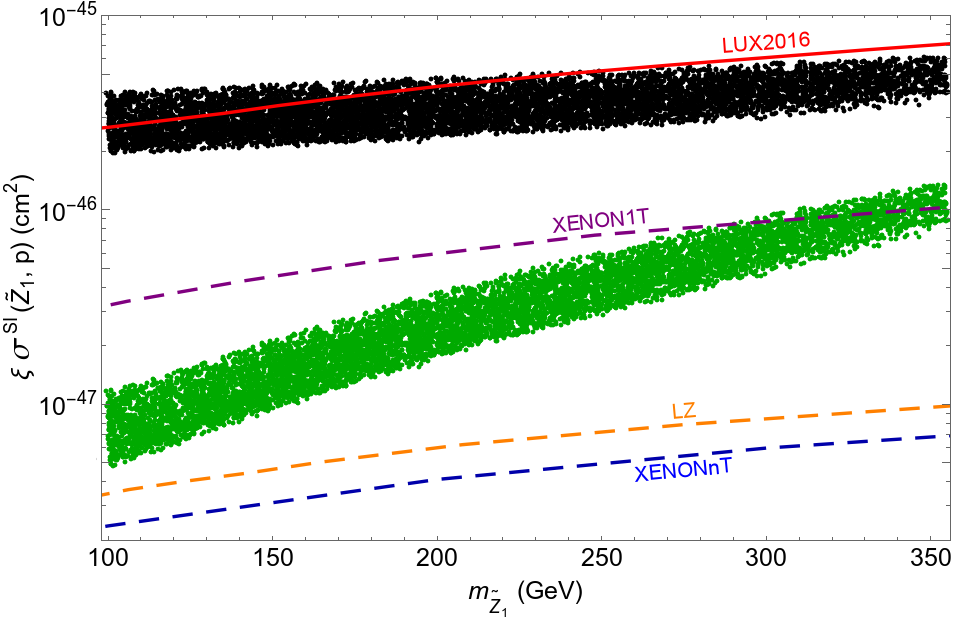}}
\subfigure[SD direct detection rate]
{\includegraphics[width=0.7\textwidth]{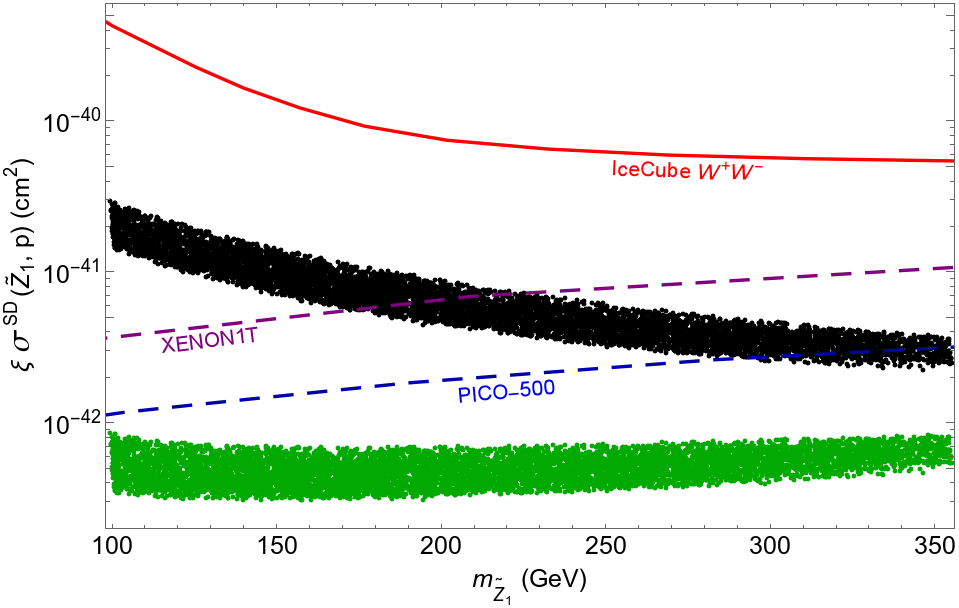}}
\caption{{\it a}) The spin-independent, and {\it b}) the spin-dependent
  neutralino-nucleon direct detection rates multiplied by fractional
  dark matter abundance $\xi\equiv \Omega_{\tz_1}^{TP}h^2/0.12$ in the
  $m_{\tz_1}$ vs. $\xi \sigma^{SI}(\tz_1, p )$ plane from a scan over
  $m_{3/2}$, $\alpha$ and $\mu$, with other parameters fixed as in the
  benchmark model.  The black points have $\xi =1$ while the green
  points have $\xi <1$ corresponding to the fraction given by thermally
  produced higgsinos. The current LUX bound is denoted by the solid
  line, while the projected reaches of several noble liquid direct
  detection experiments are shown by the dashed lines in frame {\it
  a}). In frame {\it b}), we show the current IceCube limit by the
  red-solid line and projected reaches of future detectors XENON1T
  (dashed-purple) and PICO-500 (dashed-blue).
\label{fig:DM}
}
\end{center}
\end{figure}
In Fig.~\ref{fig:DM}{\it a}), we show the WIMP spin-independent (SI)
direct detection rates expected from nGMM in the $m_{\tz_1}$ vs. $\xi
\sigma^{SI}(\tz_1, p )$ plane. The vertical axis includes a factor of
$\xi\equiv \Omega_{\tz_1}^{TP}h^2/0.12$ to account for the possibility of a
depleted local abundance of target WIMPs. Here, we adopt matter scalar
soft terms $\sim 5$ TeV with $a_3=5.1$ and the $m_A=2$ TeV and then scan
over $m_{3/2}:10-200$ TeV, $\alpha:0-20$ and $\mu : 100-400$ GeV. We
show only the points with $\Delta_{\rm EW}<30$. The upper black points
assume that higgsinos produced by an additional non-thermal $\tz_1$
production saturate the observed dark matter density, so $\xi =1$, while
for the lower green points we assume the higgsino abundance is given by
its thermal value so that the bulk of dark matter is axions. Non-thermal
production of higgsinos from axino/saxion decays would increase
$\Omega_{\tz_1}h^2$ resulting in an increase to $\xi$ of the green points. 
Of course, the density of
neutralinos could be diluted if there was additional entropy production~\cite{Bae:2013qr}
during the history of the Universe. 
The current reach of the LUX experiment~\cite{lux} is shown as red-solid while the 
XENON1T reach~\cite{xen1t} is purple-dashed.
We see that the current LUX experiment has just started to probe the parameter space 
with $\xi=1$ while all of this space will be probed by XENON1T. 
Multi-ton noble liquid detectors such as LZ~\cite{Akerib:2015cja}, 
XENONnT (20tY exposure)~\cite{xen1t}, DarkSide-20K~\cite{Agnes:2015lwe}, 
DEAP-50T~\cite{Amaudruz:2014nsa} and DARWIN~\cite{Aalbers:2016jon}
will be required to probe the 
entire parameter space with $\xi <1$. We note these detection rates are
lower than expected from natural NUHM2 models~\cite{bbm,wimp} since both
$\xi$ is reduced and also with heavier electroweak-ino masses, the LSP
is more pure higgsino-like in nGMM. In this case, the Higgs exchange
amplitude, which depends on a product of higgsino times gaugino
couplings, is reduced in nGMM compared to NUHM2.

In Fig.~\ref{fig:DM}{\it b}), we show the spin-dependent cross sections
for the same scan as in frame {\it a}) with $\xi=1$ and $\xi<1$ (fixed
by the thermal abundance of higgsinos), along with the current bound
from the IceCube experiment (red solid line)~\cite{icecube} and
projected reaches of the XENON1T (dashed purple line) and PICO500
(dashed-blue line)~\cite{pico}. We see that the nGMM points,
even with $\xi=1$, satisfy all current bounds. This situation is quite
different from the case of the well-tempered neutralino where the
higgsino-rich neutralino branch is solidly excluded by the IceCube
data. The reason is that though higgsinos couple with full gauge
strength to the $Z$, in the case of the (nearly) pure higgsino-LSP of
the nGMM, the coupling of $Z$ to {\it identical} neutralinos (which
determines the SD cross section) vanishes when the $\tz_i \simeq
{{\tilde{h}_u \pm \tilde{h}_d} \over\sqrt{2}}$. 
We see that the XENON1T experiment will detect a signal even via spin-dependent scattering for
$m_{\tz_1}\lesssim 200$~GeV if neutralinos make up all the local DM.
Experiments like PICO-500 will be needed to probe yet higher mass values. 
Finally, we remark that if the neutralino density is determined
by its thermal value, it will escape detection via SD neutralino-nucleon
scattering in the case of the nGMM.

\begin{figure}[tbp]
\begin{center}
\includegraphics[width=0.7\textwidth]{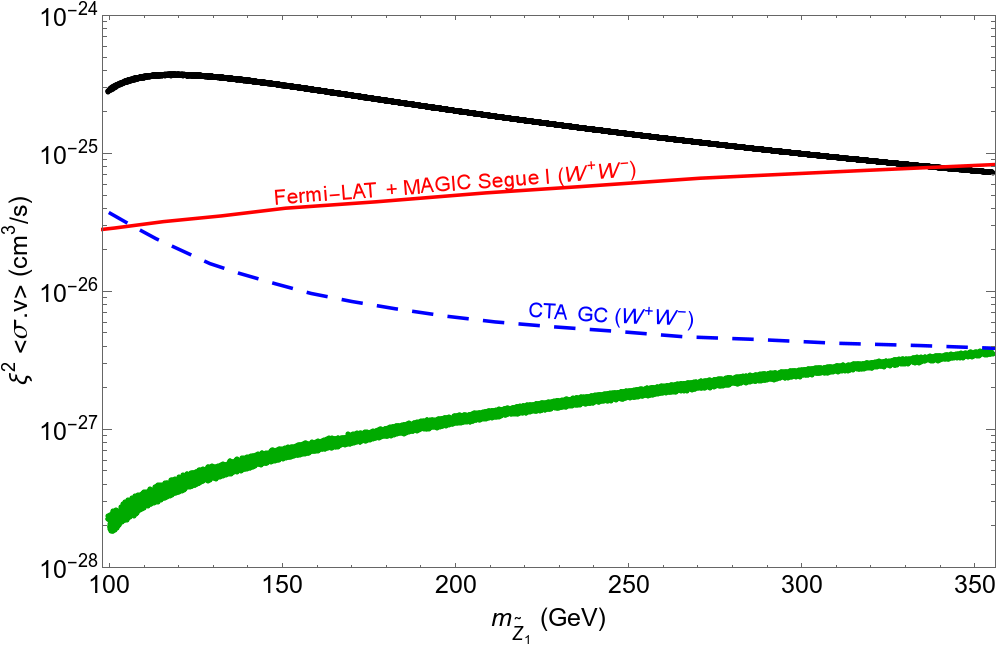}
\caption{The neutralino annihilation cross section times velocity scaled
  by $\xi^2$ 
  in the $m_{\tz_1}$ vs. $\xi^2\langle \sigma v\rangle$ plane from a scan
  over $m_{3/2}$, $\alpha$ and $\mu$. 
  The black points have $\xi =1$ while the green
  points have $\xi <1$
  . The solid red-line shows the upper bound from
  FERMI-LAT collaboration (assuming neutralino annihilation to $W^+W^-$
  pairs) while the dashed blue line shows the
  corresponding projected reach of CTA.
\label{fig:fermi}
}
\end{center}
\end{figure}
In Fig.~\ref{fig:fermi} we plot the values of $\xi^2\langle \sigma v\rangle$, 
the thermally-averaged neutralino annihilation cross section times velocity, 
versus the lightest neutralino mass for the same scan as in Fig.~\ref{fig:DM}. 
Higgsino-like neutralinos in the range of interest dominantly annihilate to $W^+W^-$ pairs. 
As before, we show results for $\xi = 1$ by black dots, and for
$\xi$ determined assuming the neutralino relic density is given by its
thermal value by green dots. The solid, red line shows the upper bound on
the neutralino cross section, assuming annihilation to $W$ boson pairs,
obtained in Ref.~\cite{fermi} by combining the dwarf-spheroidal data
from the Fermi-LAT collaboration and the MAGIC
collaboration.\footnote{For the mass range of our interest the limit is
mainly dominated by FERMI-LAT observations.} Taken at face value, this
analysis excludes the possibility that higgsino relics dominate the CDM
density over almost the entire mass range favoured by electroweak
naturalness.\footnote{For the nGMM model scan that we are discussing, 
we have checked that the Fermi-LAT constraint restricts 
the higgsino component of the dark matter
to no more than $\sim$35\% (50\%) [85\%] for $m_{\tz_1} = 150$~GeV
(200~GeV) [300~GeV].}
In contrast, if we assume that the higgsino contribution to
the DM density is given by its thermal expectation, it appears that in
nGMM dark matter indirect detection via the gamma ray signal would be
very difficult even at the proposed ground-based Cherenkov Telescope
Array, projections for which are shown by the dashed blue line in the
figure~\cite{cta}.

\section{Concluding Remarks}

The simplest renditions of the very intriguing model of mirage mediation
seem to be strongly disfavoured by naturalness considerations, when
combined with the measured value of the Higgs boson mass and lower
limits from the LHC on superparticle masses.  However, several groups
have observed that while MM gaugino mass predictions are very robust,
the scalar sector is quite sensitive to the mechanisms for moduli
stabilization and potential uplifting.  Here, we advocated a generalized
version of MM where discrete parameters depending on modular weights are
elevated to continuous ones to parametrize more general possibilities
for moduli stabilization and potential uplifting.  The added flexibility
of general mirage mediation allows for construction of natural GMM
models which are consistent with LHC Higgs mass measurements and
sparticle search constraints.  We exhibit a benchmark point with a
natural superpartner spectrum which maintains mirage unification in the
gaugino sector.  The resulting spectrum, while highly natural, will
likely elude LHC searches even at very high luminosity.  In the nGMM,
prospects for dark matter detection are also modified significantly from
expectations in natural scenarios with GUT scale gaugino mass
unification.  The possibility that (non-thermally produced) higgsinos
comprise all the DM appears to be excluded by the combined
FERMI-LAT-MAGIC analysis.  If instead the WIMP density is given by its
thermal value, with the remainder being composed for instance of axions,
then multi-ton noble liquid detectors such as LZ or XENONnT or others
will be required for detection.  For the nGMM scenario, the resolving
power of ILC may well offer the best hope to unearth the predicted 
light higgsinos signal.
If ILC finds such a signal, it is possible
that fits to the gaugino masses may allow for measurements of the
relative moduli/anomaly mixing ($\alpha$) parameter and the gravitino mass $m_{3/2}$.

{\it Note added:} This revised version corrects a numerical error in 
the published version wherein a benchmark input value is corrected 
to be $a_3=5.1$. 

\section*{Acknowledgments}

We thank Kiwoon Choi and Jenny List for e-discussions.
This work was supported in part by the US Department of Energy, Office of High
Energy Physics. 
The computing for this project was 
performed at the OU Supercomputing Center for Education \& Research (OSCER) at the 
University of Oklahoma (OU).

%

%
\end{document}